\documentclass[aps,prl,reprint,superscriptaddress]{revtex4-1}

\usepackage{hyperref}
\usepackage{amsmath}
\usepackage{amsfonts}
\usepackage{amssymb}
\usepackage{graphicx}

\begin{document}


\title{Evidence of surface loss as ubiquitous limiting damping mechanism in SiN micro- and nanomechanical resonators}



\author{L. G. Villanueva}
\affiliation{Advanced NEMS Group, {\'E}cole Polytechnique F{\'e}d{\'e}rale de Lausanne (EPFL), Lausanne, Switzerland}
\author{S. Schmid}
\email[]{sils@nanotech.dtu.dk}
\affiliation{Department of Micro- and Nanotechnology, Technical University of Denmark, DTU Nanotech, DK-2800 Kgs. Lyngby, Denmark}


\date{\today}

\begin{abstract}
Silicon nitride (SiN) micro- and nanomechanical resonators have attracted a lot of attention in various research fields due to their exceptionally high quality factors ($Q$s). Despite their popularity, the origin of the limiting loss mechanisms in these structures has remained controversial. In this paper we propose an analytical model combining acoustic radiation loss with intrinsic loss. The model accurately predicts the resulting mode-dependent $Q$s of a low-stress silicon-rich and a high-stress stoichiometric SiN membrane. The large acoustic mismatch of the low-stress membrane to the substrate seems to minimize radiation loss and $Q$s of higher modes ($n \wedge m \geq 3$) are limited by intrinsic losses. The study of these intrinsic losses in low-stress membranes with varying lengths $L$ and thicknesses $h$ reveals an inverse linear dependence of the intrinsic loss with $h$ for thin resonators independent of $L$. This finding was confirmed by comparing the intrinsic dissipation of arbitrary (membranes, strings, and cantilevers) SiN resonators extracted from literature, suggesting surface loss as ubiquitous damping mechanism in thin SiN resonators with $Q_{surf} = \beta \cdot h$ and $\beta = 6\times10^{10} \pm4\times 10^{10}$~m$^{-1}$. Based on the intrinsic loss the maximal achievable $Q$s and $Q\cdot f$ products for SiN membranes and strings are outlined.
\end{abstract}


\maketitle


Since the discovery of the exceptionally high quality factors ($Q$) of nanomechanical silicon nitride (SiN) resonators \cite{verbridge2006high,zwickl2008high}, SiN strings and membranes have become the centerpiece of many experiments in the fields of cavity optomechanics \cite{thompson2008strong,Gavartin2012,Purdy2013,Brawley2014,wilson2009cavity,Faust2012,Anetsberger2010,Hammerer2009,Camerer2011,Andrews2014,Bagci2014} and sensor technology \cite{Hanay2012,Barton2010,Yamada2013,Schmid2013,Larsen2013,Schmid2014d}. For example in cavity optomechanics a high $Q$ at high frequencies is required in order to advance towards the quantum regime of the mechanical resonators, and in resonant sensors a high $Q$ enables a better resolution. Despite the continuous effort to understand and optimize $Q$ of SiN resonators, the underlying source of the limiting mechanism has remained controversial. On the one hand it has been suggested by several groups that SiN resonators are limited by intrinsic losses \cite{Schmid2011,unterreithmeier2010damping,Adiga2012}. On the other hand it has recently been suggested that radiation loss is the limiting factor for $Q$ in SiN membranes \cite{Chakram2014}. In this paper we show that a model which combines intrinsic and acoustic radiation losses accurately predicts the mode-dependent $Q$s of low- and high-stress SiN membranes. Finally, we show that the intrinsic loss in thin arbitrary SiN resonators scales with thickness. This is evidence that surface loss is the ubiquitous limiting damping mechanism in micro- and nanomechanical SiN resonators, such as strings, membranes, and cantilevers.

The exceptionally high $Q$s of SiN resonators originate from the high intrinsic tensile stress $\sigma$ which increases the stored energy without significantly increasing the energy loss during vibration \cite{unterreithmeier2010damping,Schmid2011,Yu2012}. Assuming the energy loss to be coupled to the local out-of-plane bending during vibration, the intrinsic quality factor of a square membrane under tensile stress $Q_{intr,\sigma}$ is given by \cite{Yu2012}
\begin{equation}\label{eq:Qsigma}
Q_{intr,\sigma} \approx Q_{intr}\cdot \left[2\lambda + (n^2+m^2)\pi^2 \lambda^2 \right]^{-1}
\end{equation}
with
\begin{equation}
\lambda = \sqrt{\frac{1}{12}\frac{E}{\sigma}\frac{h^2}{L^2}}
\end{equation}
where $Q_{intr}$ is the intrinsic quality factor of the relaxed resonator without the tensile stress (like for example a cantilever), $n,m$ are the mode numbers, $E$ the Young's modulus, $h$ the thickness, and $L$ is the side length. The expression for strings can also be developed and the final result is (\ref{eq:Qsigma}) with $m=0$ and $n$ as the mode number, which is equal to an earlier model for $Q$ of loaded wires \cite{Gonzfilez1994}. The value in square brackets in (\ref{eq:Qsigma}) is a $Q$-enhancing factor that comprises two terms. The left term is independent of the mode number and comes from the local curvature of the resonator at the clamped ends. The right term is dependent on the mode numbers and originates from the curvature of the anti-nodes. As per definition of a string or membrane $\lambda\ll1$ \cite{boisen2011cantilever}. Hence, the left term is a lot larger, that is, the damping due to the membrane curvature at the clamped ends usually dominates $Q_{intr,\sigma}$. The local bending at the clamping is decreasing exponentially with a decay length $L_c = L\lambda$ \cite{Gonzfilez1994,Cagnoli2000,Yu2012}. For stoichiometric SiN $L_c \approx 5\times h$ and the peak intrinsic damping for a 30~nm thick resonator thus happens within a 150~nm wide band at the resonator ends close to the clamping.

Besides the intrinsic energy loss, the resonators can lose energy through phonons tunneling into the substrate, so-called acoustic radiation loss. It has been suggested that acoustic radiation loss in SiN membranes is strongly mode dependent and that modes with low mode numbers typically are limited by radiation loss \cite{Wilson-Rae2008}. An analytical model based on the coupling of membrane modes to free modes of the substrate has been fully developed \cite{Wilson-Rae2008,wilson2011high}. For the sake of simplicity, we provide here the asymptotic limit for a square membrane \cite{Wilson-Rae-asym}
\begin{equation}\label{eq:Qc}
Q_{rad} \approx   1.5 \alpha \frac{\rho_s}{\rho_r}\eta^3 \frac{n^2 m^2}{(n^2 + m^2)^{3/2}} \frac{L}{h}
\end{equation}
with the "acoustic mismatch" (phase velocity ratio) between a semi-infinite substrate and the resonator
\begin{equation}\label{eq:eta}
\eta \approx \sqrt{\frac{E_s}{\sigma}\frac{\rho_r}{\rho_s}},
\end{equation}
with the mass densities $\rho_s$ and $\rho_r$ of the substrate and resonator, respectively, and the Young's modulus of the substrate $E_s$. The pre-factor $\alpha$ is a fitting parameter correcting for substrate imperfections resulting from the specific chip mounting conditions. Under ideal conditions of a semi-infinite substrate $\alpha = 1$. Eq.~(\ref{eq:Qc}) is valid under the condition $n,m \gg \sqrt{n^2+m^2}/\eta$. Typically, $\eta \gg 1$ for SiN membranes and thus the radiation loss model is valid for all $n \sim m$. 
Destructive interference of the waves radiating into the substrate can lead to a suppression of the acoustic radiation loss for increasing harmonic modes ($n=m$) \cite{Wilson-Rae2008}. From (\ref{eq:Qc}) it can be seen that acoustic radiation loss is minimal for harmonic modes $n = m$ and the envelope of maximal values is increasing linearly with the mode numbers $Q_{rad} \propto n$. For strings, $Q_{rad} \propto L/w$ is predicted to be a function of the string width $w$ \cite{cross2001elastic}. This effect has been observed with SiN strings where $Q$ increased with decreasing width and approached an asymptotic limit given by intrinsic losses \cite{Schmid2011}.


\begin{figure}
  \includegraphics[width=0.44\textwidth]{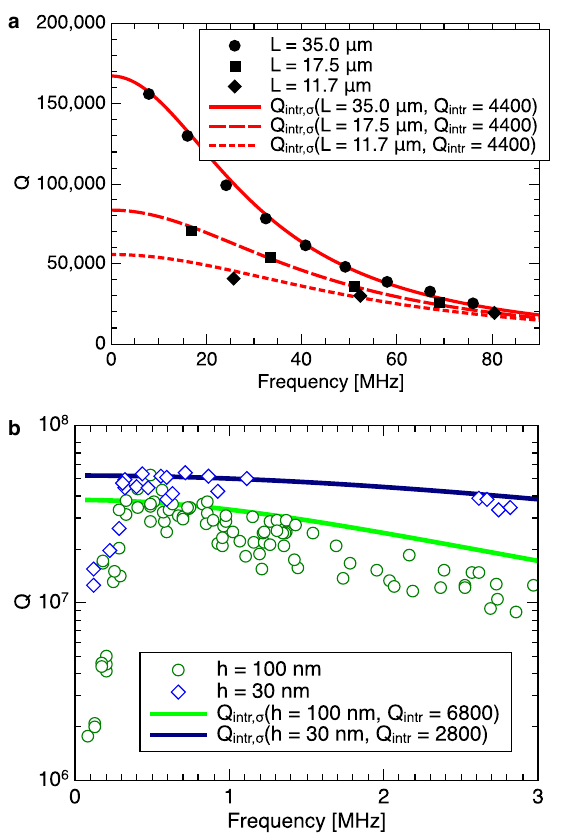}
  \caption{(a) $Q$ values for increasing flexural modes of nanomechanical stoichiometric SiN strings with varying lengths ($\sigma=942$~MPa, $h=100$~nm) \cite{unterreithmeier2010damping}. All quality factors for the strings with different length are fitted with $Q_{intr,\sigma}$ with a single $Q_{intr}=4400$. (b) $Q$ values for increasing flexural mode numbers of 5~mm$\times$5~mm stoichiometric SiN membranes with two different thicknesses ($\sigma=994$~MPa, determined from the resonance frequency for $\rho = 3000$~kg/m$^3$) \cite{Chakram2014}. The solid lines represent $Q_{intr,\sigma}$.}
  \label{fig:1}
\end{figure}

First, examples of $Q$ values of stoichiometric SiN strings and membranes from literature are discussed with respect to the intrinsic (\ref{eq:Qsigma}) and acoustic radiation loss (\ref{eq:Qc}). In Fig.~\ref{fig:1}a, the intrinsic damping model (\ref{eq:Qsigma}) is applied to $Q$ values from nanomechanical SiN string resonators \cite{unterreithmeier2010damping}. The intrinsic $Q$ model accurately predicts the measured $Q$s of several modes of 3 strings with different lengths for a common $Q_{intr} = 4400$. Apparently, these strings are narrow enough such that acoustic radiation losses can be neglected and the measured $Q$s are clearly limited by intrinsic damping, thereby confirming the conclusions made by the authors of the original paper \cite{unterreithmeier2010damping,Faust2014}. The $\lambda > 0.01$ of these strings is relatively large and the influence of the anti-nodal bending becomes significant which can be seen in the steep $Q$ decrease with higher modes. 

In Fig.~\ref{fig:1}b, the intrinsic model is applied to the measured maximal $Q$s of two SiN membrane resonators with a different thickness $h$ \cite{Chakram2014}. There are two regimes: an initial increase of $Q$ with frequency (below $\sim 300$~kHz) and a maximal plateau with a $Q$-envelope that slightly decreases with higher frequencies (above $\sim 300$~kHz). In the former regime, $Q$ is increasing with frequency and it has repeatedly been shown that $Q$s of these lower modes can be increased by minimizing the contact between chip and support \cite{wilson2009cavity,Wilson2012,Chakram2014}. Thereby $Q$ could be lifted up to the maximal $Q$ plateau, while the magnitude of the plateau remained independent of the mounting condition (see Supplementary Information) \cite{wilson2009cavity,Wilson2012}. A similar effect has been observed with SiN strings where the free suspension of the chip suppressed the string width-dependent radiation losses (see Supplementary Information) \cite{Schmid2011}. It can thus be concluded that radiation loss is responsible for the low $Q$s at low frequencies. In the latter plateau regime on the other hand, the observed slight downwards trend is accurately predicted by the intrinsic damping model. With a minute $\lambda < 9.4\times 10^{-5}$ the additional intrinsic loss from the anti-nodal bending of the membrane is low and the maximal $Q$ plateau remains relatively high also at higher modes. In contrast, the radiation loss model (\ref{eq:Qc}) predicts a $Q$ envelope that is linearly increasing with frequency, which is not the case. The accurate congruence of the experimental $Q$ data for higher modes with the intrinsic model (\ref{eq:Qsigma}) strongly suggests that the maximal $Q$ values of higher modes are ultimately limited by intrinsic losses. This evidence is contradicting the conclusions made by the authors of the original paper who argued that radiation loss is the only limiting damping mechanism in both regimes \cite{Chakram2014}.

According to (\ref{eq:Qc}), $Q_{rad}$ is a function of the acoustic mismatch $\eta$ between the resonator and a semi-infinite substrate. At low frequencies however, the resulting wavelengths in the substrate can become larger than the Si chip. In this case the chip-mount (including  chip holder, glue, tape, piezo-shaker, etc) has to be considered part of the substrate. Besides the higher radiation loss at lower modes according to (\ref{eq:Qc}), the higher sensitivity of the lower frequency $Q$s in Fig.~\ref{fig:1}b to the chip mounting conditions can be explained by a reduced $\eta$ due to the longer wavelengths (an approximate model taking into account the wavelength-dependent $\eta$ is presented in the Supplementary Information). Accordingly, it has recently been demonstrated that $Q$ of nanomechanical SiN strings deteriorates when the acoustic mismatch is reduced by touching the anchor area with an AFM tip \cite{Rieger2014}. A successful way of suppressing radiation losses is to locate the mechanical structure within a well designed phononic bandgap structure. This removes the free frame modes around the membrane and suppresses the probability of phonon tunnelling, i.e. radiation loss \cite{Tsaturyan2014,Yu2014}. The measured maximal $Q$s of modes with negligible radiation loss of such a SiN membrane had maximal $Q$ values that correspond to expected values obtained with similar membranes without the phononic bandgap. This is a strong evidence that $Q$s in SiN membranes ultimately can be limited by intrinsic losses if the chip is mounted carefully. 




\begin{figure}
  \includegraphics[width=0.44\textwidth]{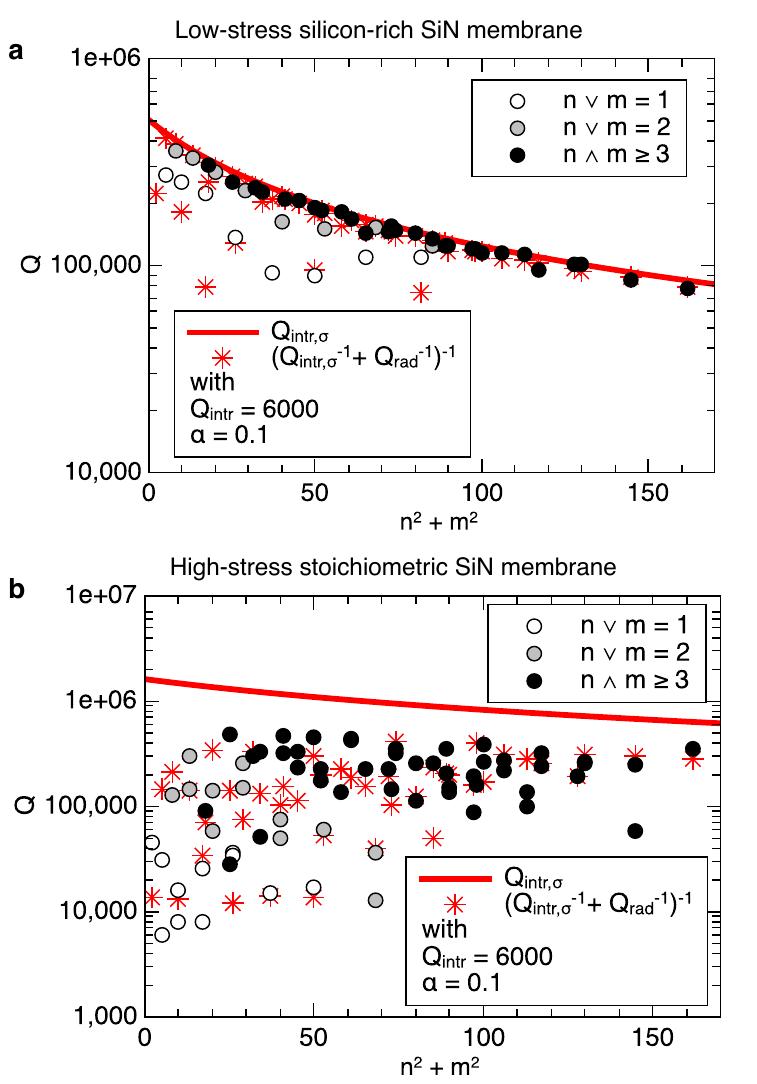}
  \caption{$Q$s for modes ($n \wedge m \leq 9$) of a a) square SR-SiN membrane ($L\approx250$~$\mu$m, $\sigma=92$~MPa, $h\approx 100$~nm) and b) square stoichiometric SiN membrane ($L\approx250$~$\mu$m, $\sigma=988$~MPa, $h\approx 100$~nm). The chips were fixed with a double sticky carbon disc (Agar Scientific). The red line represents the the highest quality factor value envelope due to intrinsic loss (\ref{eq:Qsigma}). The red stars represent the fit of all quality factor values with (\ref{eq:Q}), combining intrinsic and acoustic radiation loss based on the exact model developed by Wilson-Rae et al. \cite{wilson2011high}. The material properties of silicon were used for $E_s = 130$~GPa and $\rho_s = 2300$~kg/m$^3$.}
  \label{fig:2}
\end{figure}

This reduction of $Q$ for low mode numbers has been observed several times and it is been associated with radiation loss \cite{Yu2012,wilson2009cavity,Wilson2012,Jockel2011}. Indeed, there are strong indications that the overall mode dependence of Q is best described by a combination of both models. In order to test this we compare low-stress silicon-rich SiN (SR-SiN) (from Norcada) and high-stress stoichiometric SiN (fabricated in-house) membrane $Q$ data to a combined model that takes into account both, intrinsic and acoustic radiation losses
\begin{equation}\label{eq:Q}
Q^{-1} = Q_{intr,\sigma}^{-1} + Q_{rad}^{-1}.
\end{equation}
The membranes were characterized in the frequency domain with a lock-in amplifier (Zurich Instruments HF2PLL) in high vacuum (pressure $< 10^{-5}$~mbar) at room temperature. The membrane motion was actuated in the linear regime with a piezoelectric shaker and detected with a laser vibrometer (MSA-500 Polytec GmbH).

Figs.~\ref{fig:2}a{\&}b show the measured $Q$s for various modes of a SR-SiN and stoichiometric SiN membrane, respectively, with equal dimensions. The combined model (\ref{eq:Q}), based on the exact solution of the radiation loss model \cite{wilson2011high}, is predicting the measured values of both membranes with good accuracy for a single chosen set of parameters $Q_{intr}$ and $\alpha$. All the modes in Fig.~\ref{fig:2} fulfill the conditions required for the validity of the radiation loss model. The maximal $Q$s of the low-stress membrane Fig.~\ref{fig:2}a are producing an envelope of maximal $Q$ values which is accurately described by the intrinsic damping model (\ref{eq:Qsigma}) (red line). Hence, the maximal $Q$s of the SR-SiN membrane seem to be clearly limited by intrinsic losses. In contrast, the peak $Q$s of the high-stress membrane are below the intrinsic loss envelope and they thus seem to be limited by radiation loss. The combined model (\ref{eq:Q}) is shown as red stars. In both membranes, modes with $n \vee m \leq 2$ are suppressed strongest by acoustic radiation loss, as predicted by the model, and as it was suggested by \cite{Yu2012}. Both Si chips were fixed to the piezoelectric actuator with a double sticky carbon tape. The resonance frequencies are in the MHz-regime which results in wavelengths in the Si that are larger than the Si chip thickness. Hence, the carbon tape and the piezo-shaker become part of the substrate. The lower Young's modulus of the tape reduces the acoustic mismatch compared to a pure Si substrate, which is reflected in the fit parameter $\alpha = 0.1 < 1$.  The lower stress in the SR-SiN membrane results in a better acoustic mismatch $\eta$ and a lower $Q_{intr}$-envelope so that the maximal $Q$s are limited by intrinsic losses, which entails $Q$s that are less scattered compared to the high-stress membrane. The same effect has been observed with high- and low-stress SiN strings \cite{Schmid2011}. Also, the relatively large $\lambda$ has the consequence that the damping contribution from the anti-nodal bending becomes significant, which yields the peculiar reduction of $Q$ with higher mode numbers, as can also be seen e.g. in Fig.~\ref{fig:1}a. This distinct pattern of intrinsic damping increases the confidence in the correct model application. SR-SiN membranes are thus the optimal structures to investigate the origin of the intrinsic loss, which is presented in the following part.

\begin{figure}
  \includegraphics[width=0.44\textwidth]{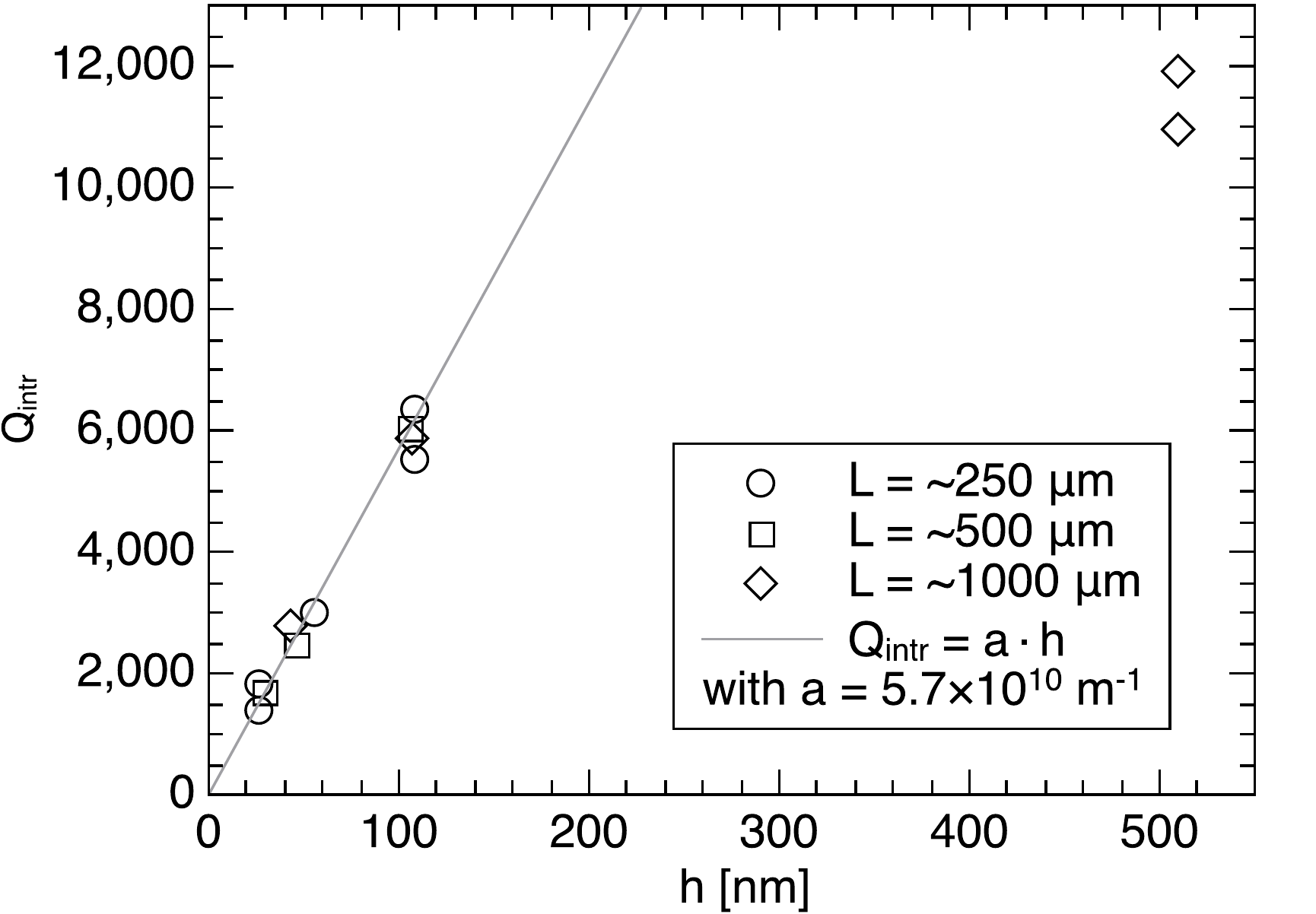}
  \caption{Intrinsic quality factors $Q_{intr}$ as a function of membrane thickness $h$ of SR-SiN membranes. The values represent the maximal values within the asymptotic $Q_{intr}$-envelope. The tensile stress varied strongly between 60~MPa~$<\sigma <$~253~MPa (determined by means of resonance frequency). $Q_{intr}$ was extracted by means of (\ref{eq:Qsigma}). The lowest values are fitted with a linear slope.}
  \label{fig:3}
\end{figure}
 
Fig.~\ref{fig:3} shows the extracted $Q_{intr}$ from the maximal $Q$ envelope given by intrinsic losses (\ref{eq:Qsigma}) from a set of square SR-SiN membranes with varying thicknesses $h$ and lengths $L$. The complete set of measured $Q_{intr}$ are plotted in the Supplementary Information. The $Q_{intr}$ values increase steadily with membrane thickness $h$, independent of the membrane size $L$. For low $h$ the increase is following a linear trend (see linear slope line). In accordance with this trend are the observed thickness dependent $Q_{intr}$s obtained from the intrinsic model in Fig.~\ref{fig:1}b. A similar linear trend has been observed with Qs of SiN micro cantilevers and was assigned to surface loss $Q_{surf}(h) = \beta \cdot h$, with a slope $\beta$ \cite{yasumura2000quality}. Hence, the observed linear relationship in Fig.~\ref{fig:3} of $Q_{intr}$ with $h$ is strong evidence of surface loss. For structures with a reduced surface to volume ratio, surface loss will become obsolete and the intrinsic loss will be dominated by volume loss $Q_{vol}$. This can be summarized by the formula
\begin{equation}\label{eq:Qi}
Q_{intr}^{-1}(h) = Q_{surf}^{-1}(h) + Q_{vol}^{-1}.
\end{equation}

\begin{figure}
  \includegraphics[width=0.46\textwidth]{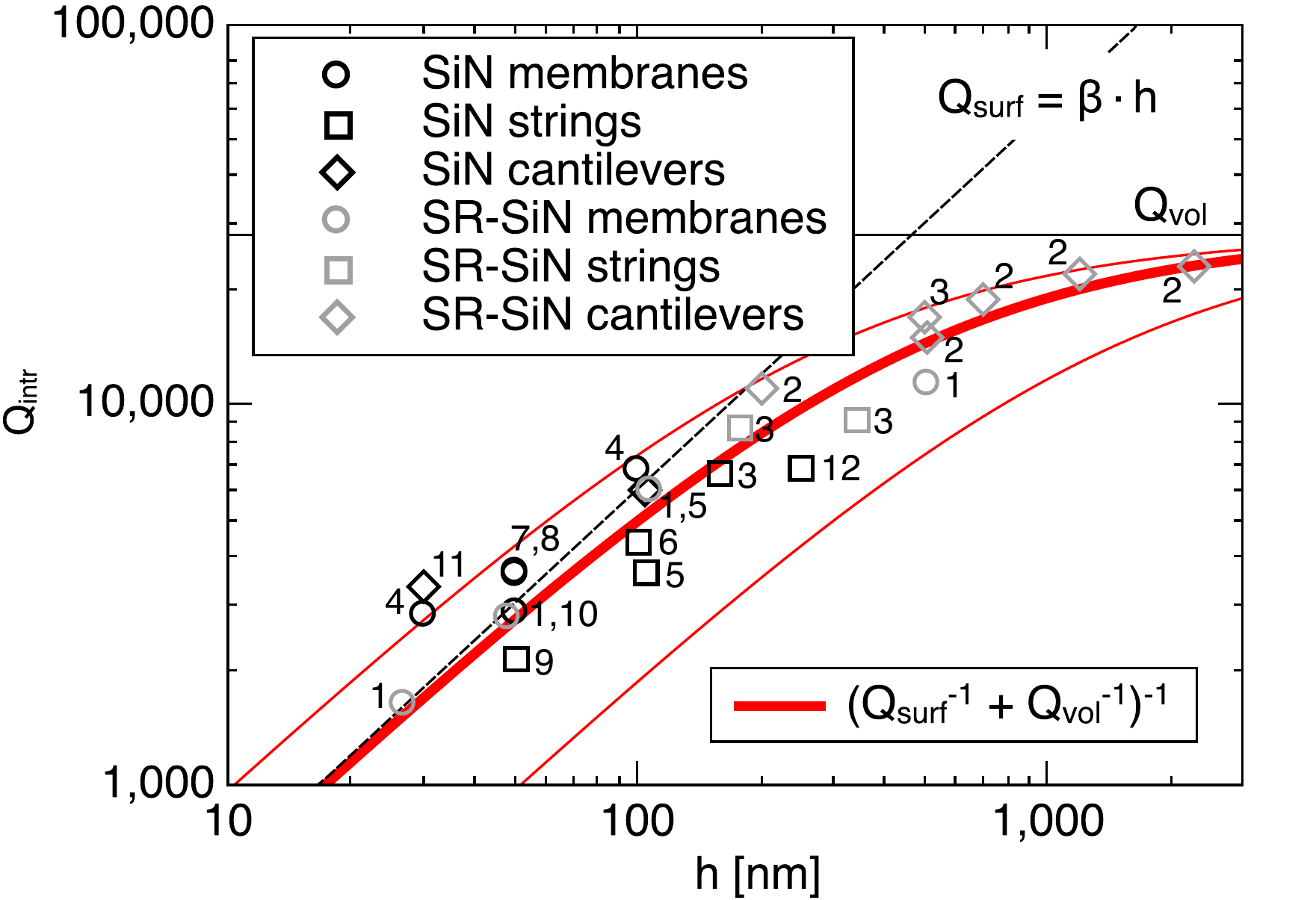}
  \caption{$Q_{intr}$ values at room temperature extracted from literature and this work as a function of structure thickness $h$. The red line represents (\ref{eq:Qi}) fitted to all values with $\beta = 6\times 10^{10}\pm4\times 10^{10}$~m$^{-1}$ and a volume loss related $Q_{vol} = 28000\pm2000$. The fine red lines represent the estimated error of $\beta$ of $\pm60$~\%. The values are taken from 1:this work, 2:\cite{yasumura2000quality}, 3:\cite{Schmid2011}, 4:\cite{Chakram2014}, 5:\cite{verbridge2006high}, 6:\cite{unterreithmeier2010damping}, 7:\cite{Tsaturyan2014}, 8:\cite{wilson2009cavity}, 9:\cite{Verbridge2008}, 10:\cite{Yu2012}, 11:\cite{southworth2009stress}, 12:\cite{Suhel2012}. All values were extracted assuming $E =240$~GPa and $\rho=3000$~kg/m$^3$. }
  \label{fig:4}
\end{figure}

In order to get more data to test the model (\ref{eq:Qi}), we extract $Q_{intr}$ values for diverse SiN resonators from literature. The values are obtained directly from maximal $Q$s of un-stressed cantilevers, and calculated by means of (\ref{eq:Qsigma}) from pre-stressed strings and membranes. All $Q_{intr}$ values are listed in Fig.~\ref{fig:4} together with the average values from Fig.~\ref{fig:3}. All values are fitted with (\ref{eq:Qi}). Apparently, the trend of all $Q_{intr}$s of all different SiN structures is described accurately by a combination of surface and volume loss. Our membranes had relatively large variations in $h$, $L$, and $\sigma$ of $\pm15$~\%, $\pm25$~\%, and $\pm75$~\%, with respect to their nominal values, which propagates to a total uncertainty in the extraction of $Q_{intr}$ of $\pm60$~\%. We took this as our error estimation for all values (thin red lines). From the fit, an average surface loss parameter of $\beta = 6\times 10^{10}\pm4\times 10^{10}$~m$^{-1}$ and a volume loss related $Q_{vol} = 28000\pm2000$ can be extracted. It seems that all different structure types made from either SR-SiN or stoichiometric SiN are ultimately limited by surface loss. Volume loss starts to significantly contribute in thicker resonators. 

The origin of the observed surface loss could be manifold, e.g. surface impurities or surface roughness. The chemical analysis with XPS (X-ray photoelectron spectroscopy) of the surface of two SiN membranes (one commercial stoichiometric LPCVD SiN membrane from Norcada, and one stoichiometric LPCVD SiN membrane fabricated in-house) revealed a high concentration of oxygen and carbon (see Supplementary Information). The same finding was made earlier by Yang et al. \cite{Yang1998} who found oxygen and carbon concentration on the surface of LPCVD SiN of 22~\% and 10~\%, respectively. It has further been shown that these specific SiN surface impurities remain after cleaning with hydrofluoric acid \cite{French1997}. Surface impurities seem to be ubiquitous in LPCVD SiN films. Surface roughness of untreated LPCVD SiN has found to be in the range of 0.3 - 3~nm \cite{Yang1998,Gui1997}. Hence, surface roughness can become a significant fraction of the total SiN thin film thickness.

\begin{figure}
  \includegraphics[width=0.44\textwidth]{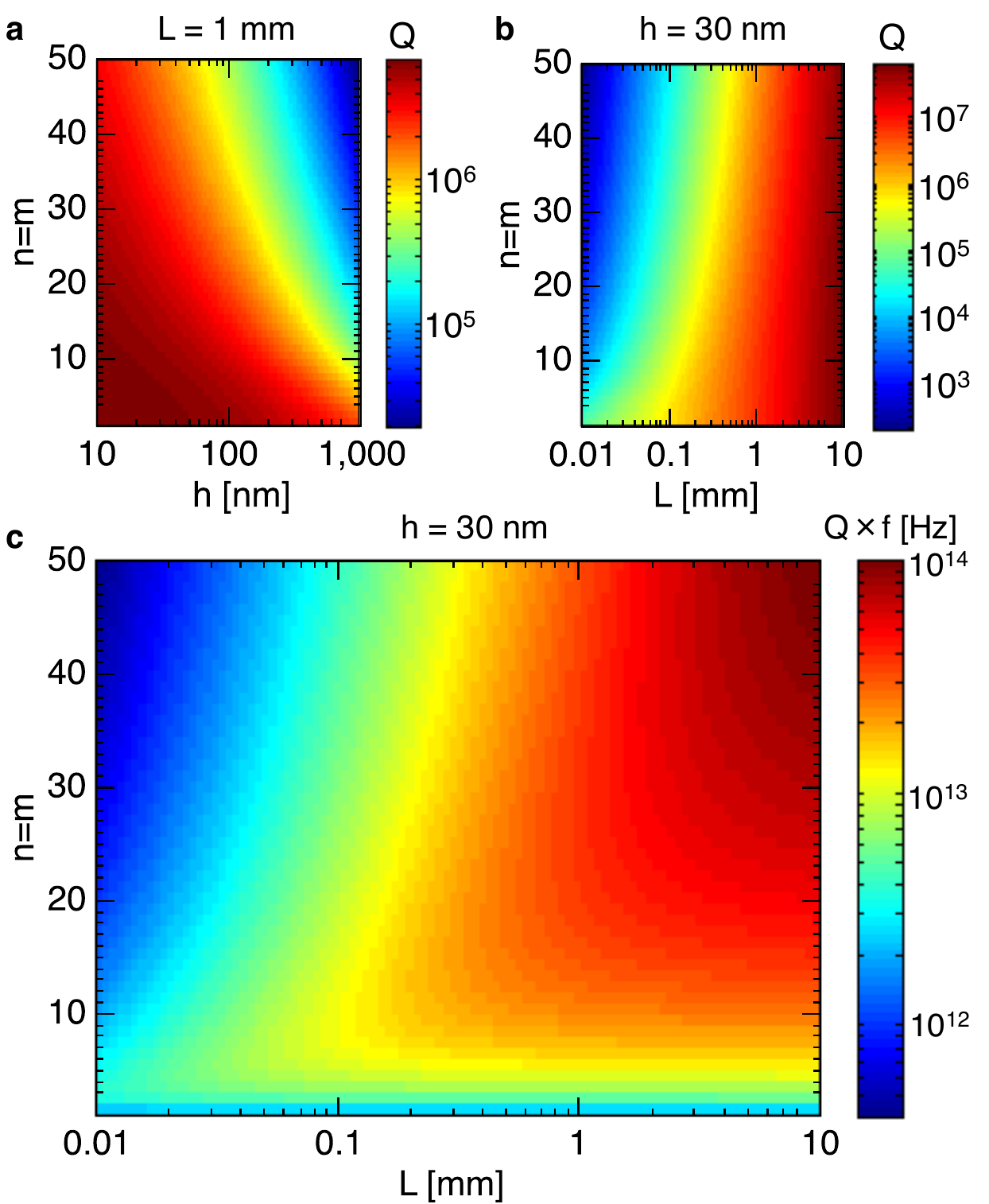}
  \caption{Prediction of maximal $Q$ and $Q \cdot f$ values obtainable with a square SiN membrane at room temperature for harmonic modes ($n = m$) that are limited by intrinsic loss. A value error of 60~\% has to be assumed.}
  \label{fig:5}
\end{figure}

Based on the $Q_{intr}$ master-curve for SiN from Fig.~\ref{fig:4}, it is now possible to predict the maximal obtainable $Q$s for harmonic modes $n=m$ of square SiN membranes that are limited by intrinsic loss. From Fig.~\ref{fig:5}a it becomes evident that the thickness does not significantly influence $Q$ of thin membranes at low mode numbers, an effect that has been observed experimentally \cite{Adiga2012}. This is a direct effect of the $Q_{intr}$ that decreases with thickness and hence counteracts the $Q$-enhancing effect of a small $h$ in (\ref{eq:Qsigma}). Thinner membranes only result in higher $Q$s at higher modes. For Fig.~\ref{fig:5}b the thickness is fixed to 30~nm. It is not surprising that larger membranes result in higher $Q$s. For large membranes, $Q$ is stable over many modes, which again can be seen in Fig.~\ref{fig:1}b. But $Q$ starts to deteriorate with mode numbers when $\lambda$ becomes large, an effect that can be seen with short SiN strings in Fig.~\ref{fig:1}a. In quantum cavity optomechanics a figure of merit is the $Q \cdot f$ product. It is a direct measure for the decoupling of the mechanical resonator from the thermal environmental bath with temperature $T$. With $Q \cdot f = k_B T/\hbar$,  $Q \cdot f > 6\times10^{12}$~Hz is the minimum requirement for room-temperature quantum optomechanics \cite{Aspelmeyera}. In that case the thermal decoherence can be neglected over one mechanical period of vibration $1/f$. The maximal $Q \cdot f$ product obtainable with a SiN membrane at room temperature is shown in Fig.~\ref{fig:5}c. It seems that the limit can not be overcome in the fundamental mode independent of membrane size, confirming the experimental findings from Wilson et al. \cite{wilson2009cavity}. For SiN string resonators, the maximal $Q$ values for low mode numbers are equal to the ones displayed in Fig.~\ref{fig:5}a{\&}b, but the $Q \cdot f$ product values have to be divided by $\sqrt{2}$. It has been shown that intrinsic damping is reduced at cryogenic temperatures which means that all predicted values in Fig.~\ref{fig:5} will increase accordingly \cite{Tsaturyan2014,Faust2014,southworth2009stress}.

In conclusion, $Q$ in pre-stressed SiN micro- and nanomechanical resonators is limited by a combination of intrinsic and acoustic radiation loss. In membranes, both respective $Q$s scale linearly with the dimensions ($L/h$). Hence, the limiting damping mechanism is mainly determined by the acoustic mismatch of the membrane to the substrate ($\eta$). In high-stress SiN membranes, $\eta$ is reduced and the maximal intrinsic loss $Q$-limit is increased, hence the resulting $Q$ values are strongly mode dependent and are scattered due to radiation loss. The maximal $Q$s can become limited by intrinsic loss by maximizing the acoustic mismatch e.g. by mounting the chip freely or with a phononic bandgap structure. In contrast, low-stress SiN membranes have a higher acoustic mismatch to the substrate and the maximal intrinsic $Q$-limit is lower. Hence, the resulting $Q$s of higher mode numbers ($n \wedge m \geq 3$) reach an upper envelope that is limited by intrinsic losses, while lower mode numbers ($n \vee m \leq 2$) can be limited by radiation loss. Generally, radiation loss is minimal for symmetric modes ($n \sim m$). In SiN strings, radiation loss scales inversely with width and narrow strings can become limited by intrinsic loss.

The intrinsic quality factors $Q_{intr}$ of thin low-stress SiN membranes scale linearly with the membrane thickness, which is strong evidence of surface loss. The same linear scaling of $Q_{surf} = \beta h$ has been confirmed by independent SiN $Q$ data taken from literature (cantilevers, strings and membranes) which is evidence that surface loss is the ubiquitous limiting damping mechanism in thin arbitrary SiN resonators with a scaling factor $\beta = 6\times10^{10} \pm4\times 10^{10}$~m$^{-1}$. For thin pre-stressed resonators that are limited by intrinsic loss, the thickness dependent surface loss is counteracting the $Q$-enhancement at low mode numbers and $Q$ can only significantly be increased with the size $L$. Finally, it seems that $Q\cdot f > 6\times 10^{12}$~Hz required for quantum cavity optomechanics at room temperature can not be reached with the fundamental mode, independent of resonator length.

\begin{acknowledgments}
The authors would like to acknowledge A. Boisen for her support, and the staff in DTU-Danchip for help in the fabrication of the membranes. The authors further thank B. Amato for the help in the laboratory, A. Schliesser for his valuable input, and I. Wilson-Rae for the generous support with the radiation loss model. This research is supported by the Villum Foundation’s Young Investigator Program (Project No. VKR023125) and the Swiss National Science Foundation (PP00P2 144695).
\end{acknowledgments}


%

\end{document}